\documentclass{article}
\usepackage{spconf,amsmath,graphicx, hyperref}
\usepackage[all]{hypcap}

\usepackage{enumitem}
\setlist{nosep, leftmargin=14pt}

\usepackage{mwe} 

\def\thickhline{\noalign{\hrule height.8pt}}
\title{PREDICTION OF CELLULAR IDENTITIES FROM TRAJECTORY AND CELL FATE INFORMATION}
\name{Baiyang Dai \textsuperscript{1}\quad Jiamin Yang \textsuperscript{2}\quad Hari Shroff \textsuperscript{3}\quad Patrick La Riviere \textsuperscript{4} \thanks{\textsuperscript{1}daibaiyang@uchicago.edu,\quad \textsuperscript{4}pjlarivi@uchicago.edu}}
\address{\textsuperscript{1} Pritzker School of Molecular Engineering (PME), University of Chicago, Chicago, IL, USA\\
\textsuperscript{2} Department of Computer Science, University of Chicago, Chicago, IL, USA\\
\textsuperscript{3} Janelia Research Campus, Howard Hughes Medical Institute (HHMI), Ashburn, VA, USA\\
\textsuperscript{4} Department of Radiology, University of Chicago, Chicago, IL, USA}
\begin{document}
%
\maketitle
\begin{abstract}
    Determining cell identities in imaging sequences is an important yet challenging task. 
    The conventional method for cell identification is via cell tracking, which is complex and can be time-consuming. 
    In this study, we propose an innovative approach to cell identification during early \textit{C. elegans} embryogenesis using machine learning. 
    Cell identification during \textit{C. elegans} embryogenesis would provide insights into neural development with implications for higher organisms including humans. 
    We employed random forest, MLP, and LSTM models, and tested cell classification accuracy on 3D time-lapse confocal datasets spanning the first $4$ hours of embryogenesis. 
    By leveraging a small number of spatial-temporal features of individual cells, including cell trajectory and cell fate information, our models achieve an accuracy of over $91\%$, even with limited data. 
    We also determine the most important feature contributions and can interpret these features in the context of biological knowledge. 
    Our research demonstrates the success of predicting cell identities in time-lapse imaging sequences directly from simple spatio-temporal features.
\end{abstract}
\begin{keywords}
    cell identification, machine learning, \textit{C. elegans} embryogenesis, classification, time-lapse imaging
\end{keywords}
\section{Introduction}
\label{sec:intro}
Cell identification is a crucial goal when imaging embryogenesis \cite{bao2006automated,wu2013spatially,moyle2021structural}, 
where it grants single-cell level analysis of cell lineage, phenotypes, and gene expression patterns. 
However, due to the low signal-to-noise ratio of the microscopy data and the sparse temporal resolution in imaging, 
identifying cells in images, even in small, stereotyped organisms like the nematode \textit{C. elegans} \cite{sulston1983embryonic}, is challenging. 
The standard approach to identify cells during development is via cell tracking \cite{bao2006automated,boyle2006acetree,mavska2023cell}. 
Generally, in cell tracking, the entire image sequence is processed, with cells detected and then traced back to the very first frame of the image sequence. 
By comparing the cell tracking result with the invariant lineage or other biological prior, cells can be identified. 
This is a complex approach, which involves several steps and can be time-consuming.

In the realm of image object classification tasks \cite{russakovsky2015imagenet}, researchers have demonstrated impressive results that surpassed human-level performance \cite{he2016deep}. 
This achievement prompts consideration for extending similar classification methodologies to directly identify cells, as we classify objects in images.
It is difficult, as the appearances of nuclei in fluorescence data are often very similar, yet classifying nuclear identity may nevertheless be possible. 
Studies have shown that cells exhibit stereotypical position and migration patterns during early \textit{C. elegans} embryogenesis \cite{Guan776062}, and the variability of cell position is highly deterministic \cite{li2019systems}. 
Here we propose a machine learning based cell classification approach to identify cells in \textit{C. elegans} embryos, based on a small number of spatial-temporal features of individual cell nuclei, including trajectory and cell fate information. 
We exploited random forest \cite{breiman2001random}, multilayer perceptron (MLP), and long short-term memory (LSTM) \cite{hochreiter1997long} as our models for cell classification and tested our models with $28$ time-lapse 3-dimensional (3D + t, hereafter referred to as `4D') confocal imaging sequences \cite{li2019systems}. 
All three models achieve outstanding performances of over $91\%$ accuracy. 
Moreover, we justified our models by examining feature contributions and linking feature importance to a cell’s naming convention in the worm embryo.

\section{Method}
\label{sec:method}

\subsection{Data}
\label{ssec:data}
\begin{figure}[ht]

    \centering
    \includegraphics[width=\columnwidth]{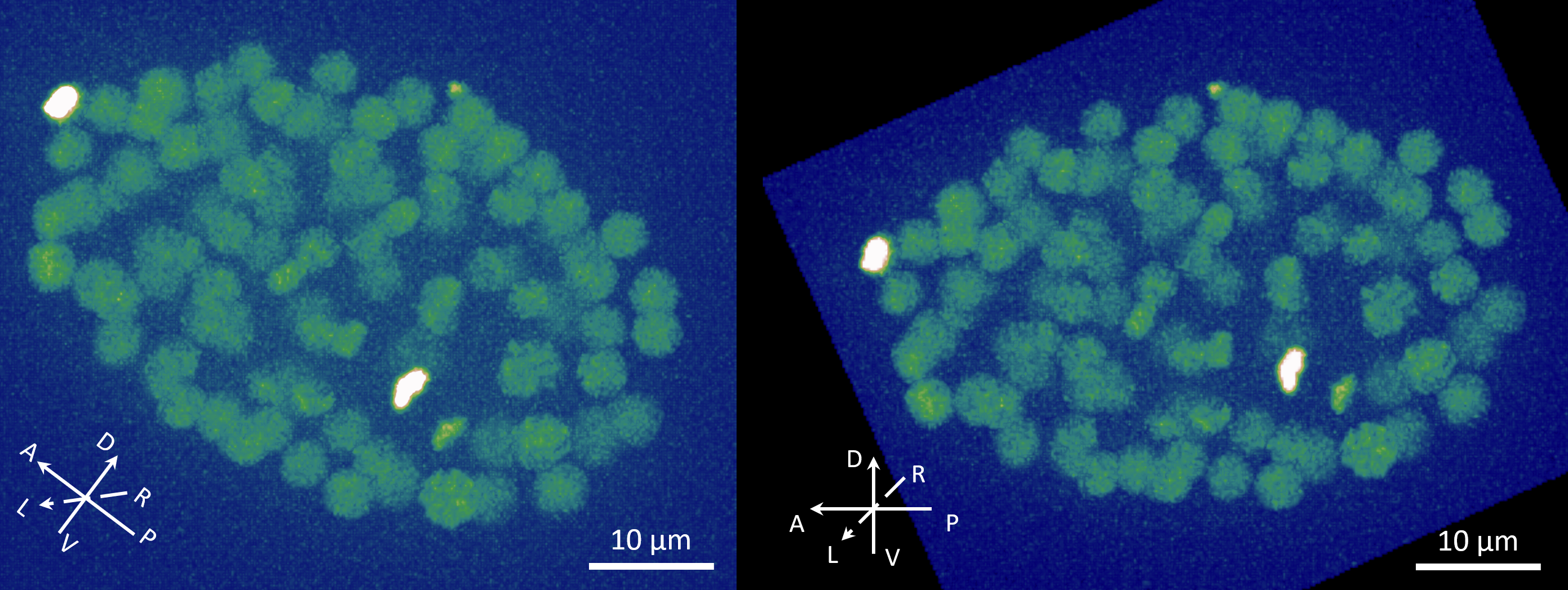}
    \caption{An example of the original embryo image (left) and the image rotated to canonical orientation (right).}
    \label{fig:image}
\end{figure}

The data used for this study comprise $28$ time-lapse 3-dimensional (4D) confocal imaging sequences for wild-type \textit{C. elegans} embryos, with a temporal resolution of $1.25$ min per frame \cite{li2019systems}. 
In the dataset, cell nuclei have already been detected and tracked until approximately the $350$-cell stage in each embryo. 
We rotated embryo images together with $(x, y, z)$ nuclear positions into canonical orientation, by assigning anterior-posterior (A-P), left-right (L-R), and dorsal-ventral (D-V) axes to X, Y, Z axes respectively. 
Fig. \ref{fig:image} shows an example of the original and the rotated image.

\subsection{Data preparation for learning}
\label{ssec:data_preparation}
We cleaned the data, and then processed and extracted features for each cell for model training and testing. 
There are $781$ unique cells in the $28$ embryos. 
We filtered out the cells that only appear in some embryos, or do not have mother or daughter cells, leaving $346$ cells remaining. 
The filtered cells are mostly very early-stage or very late-stage cells. 
Out of these $346$ cells, we further excluded $12$ cells with lifespans longer than $50$ frames, leaving us a total of $334$ unique cells for our study.

\begin{figure}[tb]

    \centering
    \includegraphics[width=\columnwidth]{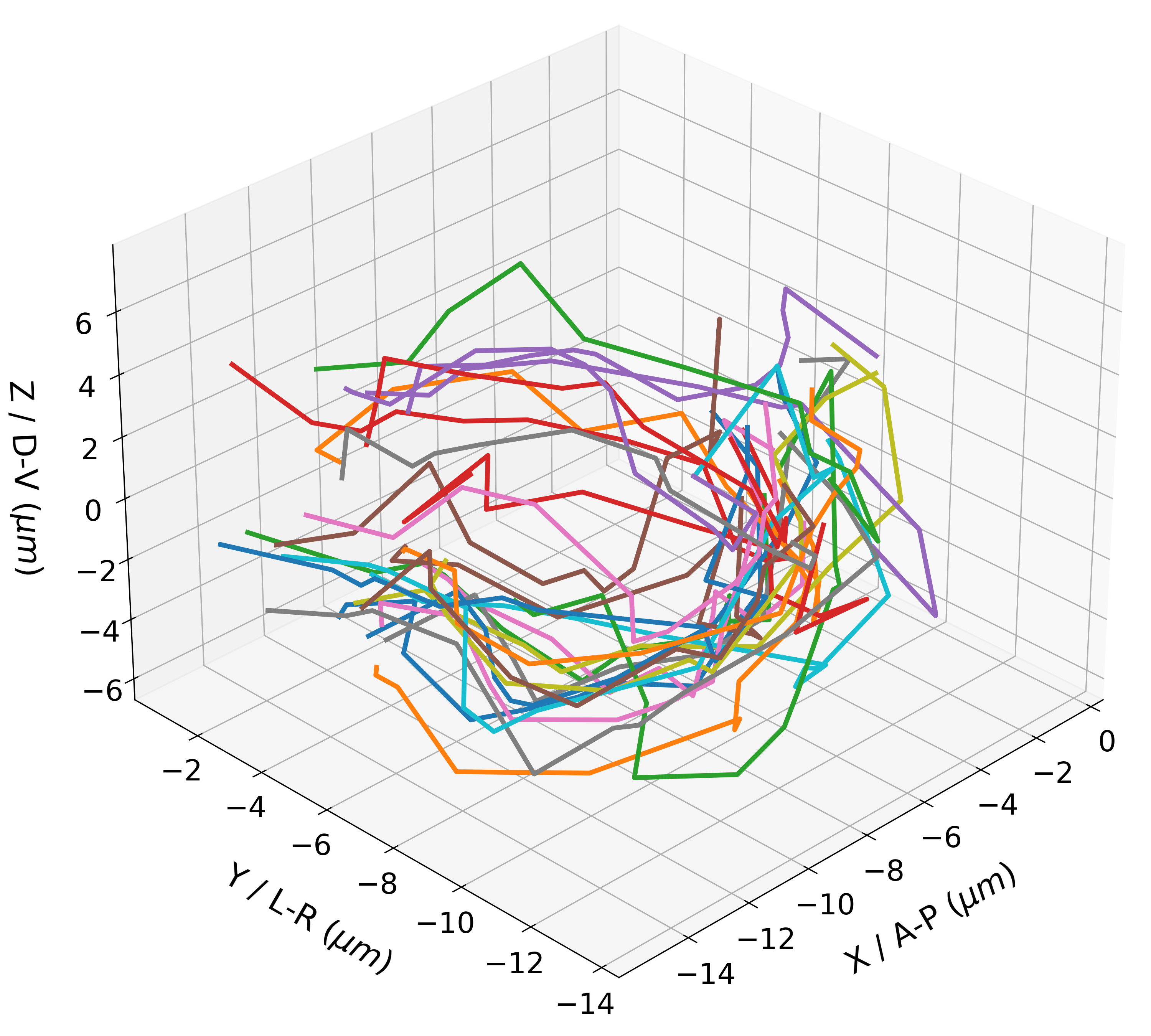}
    \caption{Trajectories (colored lines) of ABar cell in $28$ embryos.}
    \label{fig:trajectory}
\end{figure}

The features we considered for the model are some spatial-temporal features that may distinguish cells from each other, including 
cell trajectory `Traj' $(t, x, y, z)$, start time `SF', lifespan `LF', division orientation to its mother cell `DM', and the division orientations of its two daughter cells `DD'. 
We extracted and calculated these features from the cell tracking results $(cell\ label, t, x, y, z)$ associated with the images. 
For cells with lifespans shorter than $50$ frames, we pad negative $10000$ to the end of their trajectory $(t, x, y, z)$. 
The start time `SF' of each cell is its birth time, in units of frames, with respect to the end of $4$-cell stage. 
The division orientation is a 3D unit vector representing a cell’s birth location relative to its mother cell. 
Therefore, the feature dimension is $211$ (the first $200$ being trajectory features) if trajectory features `Traj' are of $(t, x, y, z)$ format, or $161$ (the first $150$ being trajectory features) if trajectory features `Traj' are of $(x, y, z)$ format. 
The trajectories of the ABar cell across all $28$ embryos are shown in Fig. \ref{fig:trajectory}.

\subsection{Models}
\label{ssec:models}
\begin{figure}[tb]

    \centering
    \includegraphics[width=\columnwidth]{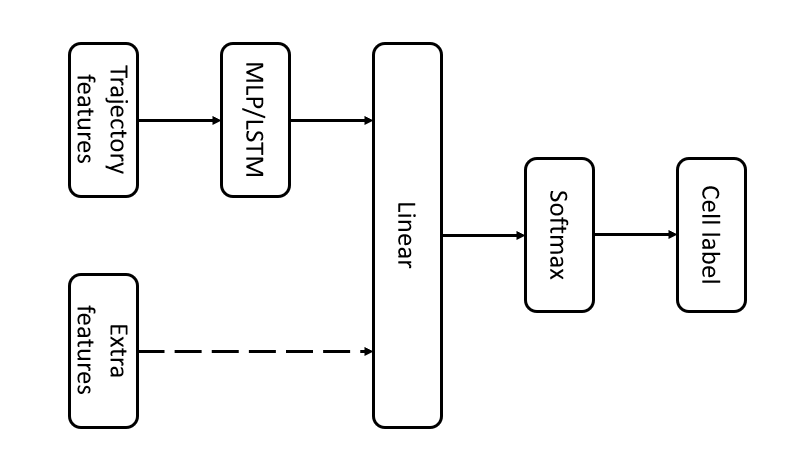}
    \caption{Model architecture for MLP and LSTM. 
    The dashed line depicts whether extra features, like start time `SF', lifespan `LF', and division orientations, are used in the model.}
    \label{fig:model}
\end{figure}

For our 4D cell classification task, we consider 3 models: random forest \cite{breiman2001random}, multilayer perceptron (MLP), and long short-term memory (LSTM) \cite{hochreiter1997long}. 
Our MLP and LSTM  models consist of two layers. 
We use a single layer MLP (with ReLU activation and residual connection \cite{he2016deep}) or single layer LSTM to learn patterns from trajectory features `Traj'. 
After learning these patterns, they are concatenated with other features and fed into a linear layer. 
The linear layer then uses a softmax function to output the predicted cell label. 
The model architecture for MLP and LSTM is illustrated in Fig. \ref{fig:model}. 
For MLP and random forest, the trajectory feature `Traj' input is $(t, x, y, z)$. 
For LSTM, we designed two models: one accepts $(x, y, z)$ format trajectory features `Traj', which we denote as LSTM; the other accepts $(t, x, y, z)$ format trajectory features `Traj', which we denote as LSTMt.

\begin{table}[!b]
    \resizebox{\columnwidth}{!}{%
    \begin{tabular}{lcccc}
    \thickhline
                        & \begin{tabular}[c]{@{}c@{}}Random\\ Forest\end{tabular} & MLP   & LSTM           & LSTMt          \\ \hline
    Traj                & 0.885                                                   & 0.813 & 0.785          & 0.817          \\
    Traj + SF           & 0.887                                                   & 0.813 & 0.827          & 0.844          \\
    Traj + SF + LF      & 0.887                                                   & 0.816 & 0.831          & 0.845          \\
    Traj + SF + LF + DM & \textbf{0.918}                                          & 0.870 & 0.899          & \textbf{0.901} \\
    Full                & \textbf{0.921}                                          & 0.879 & \textbf{0.907} & 0.898          \\ 
    \thickhline
    \end{tabular}%
    }
    \caption{Cross-validation results. Validation accuracies above $90\%$ are highlighted. Here \textit{Traj} denotes trajectory, \textit{SF} denotes start time, \textit{LF} denotes lifespan, \textit{DM} denotes division orientation to mother cell, \textit{Full} denotes all features included.}
    \label{tab:cross_validation}
\end{table}

\subsection{Cross-validation}
\label{ssec:cross_validation}
We randomly sampled $4$ embryo datasets out of 28 embryos for model testing and left the remaining $24$ embryo samples for $6$-fold cross-validation. 
We use cross-validation to find the optimal hyperparameters for the models. 
The cross-validation results are listed in Table \ref{tab:cross_validation}. 
Here we performed a feature ablation study by deleting some features, and inspected model performances. 
For the random forest model, we examined the number of trees from $10$ to $200$ in cross-validation and picked a random forest with $169$ trees according to validation accuracy. 
For MLP, LSTM, and LSTMt, we searched learning rate, weight\_decay ($L_2$ regularization), batch size, dropout rate \cite{srivastava2014dropout}, and number of hidden layers. 
We use AdamW \cite{loshchilov2018decoupled} as the optimization method to employ $L_2$ regularization, and an exponential learning rate scheduler to decay the learning rate by $0.999$ every epoch. 
All models are trained for $3000$ epochs. 
The optimal weight decay is $0.1$ for MLP, $0.15$ for LSTM, and $0.05$ for LSTMt. 
For MLP, an optimal dropout rate of $0.5$ is utilized. 
For all three models, the optimal learning rate is $0.001$ and the optimal batch size is $128$. 
All three models perform best with a single hidden layer, probably due to the small size of our dataset.

From the feature ablation study result in Table \ref{tab:cross_validation}, we can see models with full features produce the best performance. 
One notable point is that, when the feature corresponding to division orientation relative to the mother cell --- `DM' is added, the validation accuracy exhibits a large increase compared to adding other features.

\section{Results}
\label{sec:Results}

\subsection{Test accuracy}
\label{ssec:test_accuracy}

To evaluate the test performance of the models, we trained them on all $24$ cross-validation embryo samples using the optimal hyperparameters found in the cross-validation step (Section \ref{ssec:cross_validation}). 
The models were then tested on $4$ test embryo samples. 
Table \ref{tab:test_results} shows the test accuracy with feature ablations. 
It is evident that all four models achieved remarkable accuracy when using full features. 
In fact, the random forest and LSTMt models even achieved above $93\%$ accuracy. 
Furthermore, by looking at different models, we can see clearly that LSTM/LSTMt and random forest show superior performance over MLP. 
This is because LSTM excels at capturing temporal dependencies, and the small size of our dataset, which requires simple models to avoid overfitting, makes the traditional machine learning random forest model well suited to the task.

Considering that we only have $28$ embryo image sequences to work with, and there are $334$ unique cells to classify, the excellent cell classification accuracy is a surprising result. 
The simplicity of our models together with their superb performances imply that cells can be identified correctly with machine learning based classification methods based solely on spatial-temporal features of individual cells, including trajectory, start time, lifespan, and division orientations.

\begin{table}[t]
    \resizebox{\columnwidth}{!}{%
    \begin{tabular}{lcccc}
    \thickhline
                        & \begin{tabular}[c]{@{}c@{}}Random\\ Forest\end{tabular} & MLP            & LSTM           & LSTMt          \\ \hline
    Traj                & 0.897                                                   & 0.850          & 0.836          & 0.859          \\
    Traj + SF           & 0.892                                                   & 0.852          & 0.856          & 0.884          \\
    Traj + SF + LF      & 0.894                                                   & 0.844          & 0.841          & 0.882          \\
    Traj + SF + LF + DM & \textbf{0.926}                                          & \textbf{0.907} & \textbf{0.924} & \textbf{0.924} \\
    Full                & \textbf{0.932}                                          & \textbf{0.912} & \textbf{0.923} & \textbf{0.932} \\ 
    \thickhline
    \end{tabular}%
    }
    \caption{Test results. Test accuracies above $90\%$ are highlighted. Here \textit{Traj} denotes trajectory, \textit{SF} denotes start time, \textit{LF} denotes lifespan, \textit{DM} denotes division orientation to mother cell, \textit{Full} denotes all features included.}
    \label{tab:test_results}
\end{table}

\subsection{Feature importance}
\label{ssec:feature_importance}

\begin{figure}[!tb]

    \centering
    \includegraphics[width=\columnwidth]{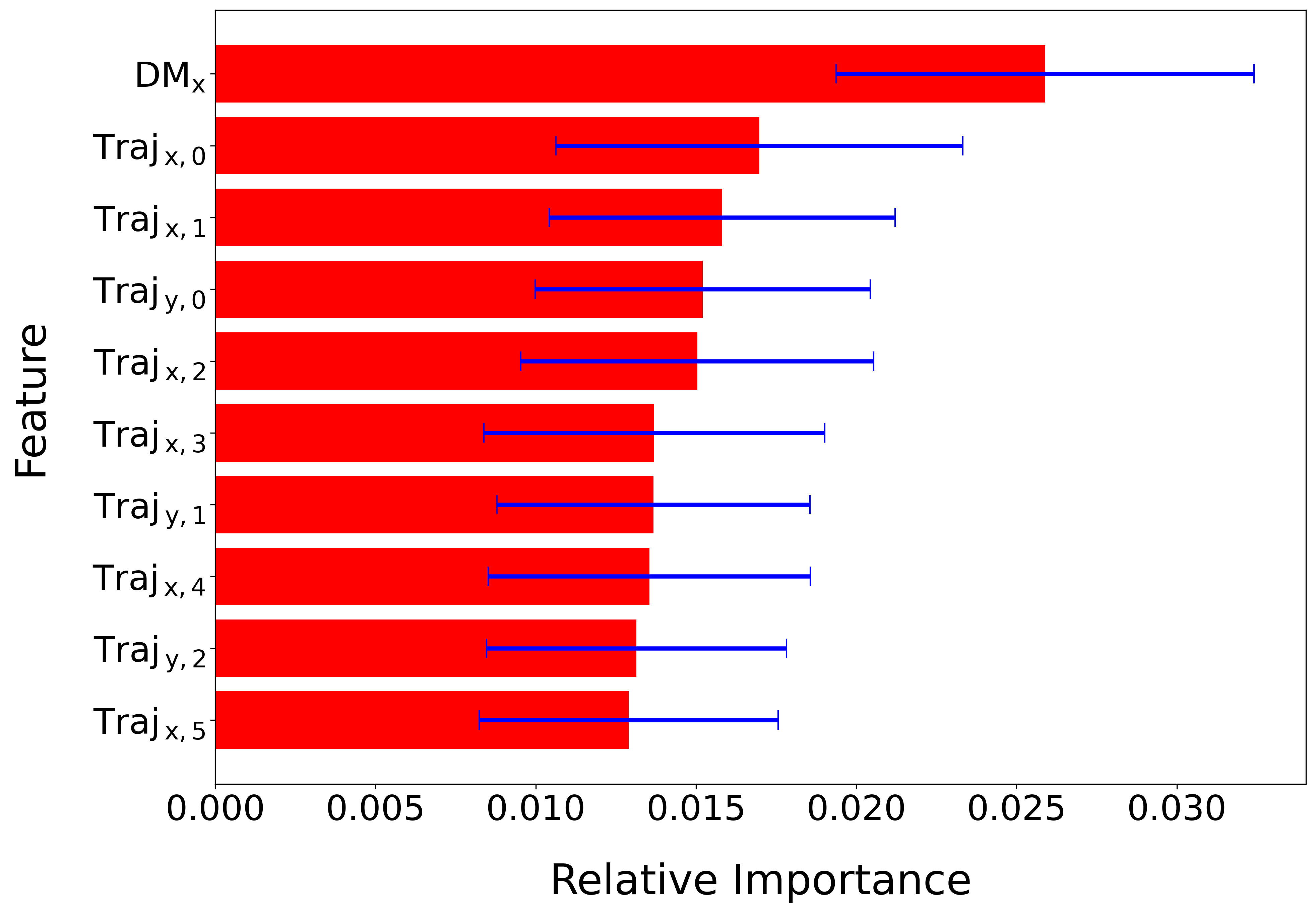}
    \caption{Top-$10$ features in random forest ranked by feature importance. 
    The blue error bars show standard deviations of feature importance for each feature. Here $\mathrm{DM_{x}}$ denotes division orientation to mother cell `DM' along X-axis, 
    $\mathrm{Traj_{\,x,0}}$ denotes cell trajectory `Traj' along X-axis at time $0$, $\mathrm{Traj_{\,y,1}}$ denotes `Traj' along Y-axis at time $1$.} 
    \label{fig:random_forest}
\end{figure}

Table \ref{tab:test_results} also illustrates the feature contributions, similar to the cross-validation result (Table \ref{tab:cross_validation}). 
From the table, we can see that the highest accuracy for each model is achieved when using all the features. 
A significant increase in accuracy occurs when adding `DM' --- a cell’s division orientation to its mother cell, and this is consistent across all the models. 
To investigate why this happens, we ranked the feature importance in random forest \cite{breiman2001random}, 
and the top-$10$ features are displayed in Fig. \ref{fig:random_forest}.
From the figure, it is obvious that feature $\mathrm{DM_{x}}$, which is cell’s division orientation to its mother cell `DM' along X-axis, is the most critical feature.
This feature indicates whether a cell is born towards the anterior or posterior (A or P) direction.
The result is reasonable as most cells in worm embryos undergo A-P divisions (these cells obtain their names by appending `a' or `p' at the end of their mother cells’ names) \cite{sulston1983embryonic}. 
Meanwhile, we can see that the remaining top-$10$ features are all trajectory features `Traj', which explains why models that use trajectory features alone, can achieve an accuracy of over $85\%$ (see Table \ref{tab:test_results}). 
This implies that it is possible to design a high-performance cell classification model by only using cell trajectories.

\section{CONCLUSIONS AND DISCUSSION}
\label{sec:conclusion}

In summary, we proposed the first machine learning based cell classification approach to identify cells in \textit{C. elegans} embryos by utilizing a cell nuclei’s spatiotemporal features. 
We demonstrated the power of our approach with four machine learning models: random forest, MLP, LSTM, and LSTMt, all of which achieved outstanding performance of over $91\%$ accuracy. 
Furthermore, we investigated the feature importance in our models and found that the cell’s division orientation to its mother cell is the most crucial feature, which accords with cell’s naming conventions in worm embryos \cite{sulston1983embryonic}. 
Moreover, our result showed that even when using trajectory features alone, cells can be classified at a high accuracy of $85\%$.

Our approach necessitates the incorporation of trajectory features, which means we need to detect and track accurately the cell of interest, to acquire its trajectory $(t, x, y, z)$. 
In contrast to cell tracking approaches \cite{bao2006automated,boyle2006acetree,mavska2023cell}, our approach only requires tracking the cell of interest and does not need to track and trace the cell all the way back to its root/ancestor cell at the very first frame of the image sequence, as in cell tracking. 
Further, our approach can output the cell name directly, whereas, in cell tracking, an additional step of comparing cell tracking results with the lineage of \textit{C. elegans} is required to get the cell name. 
Given these advantages and the simplicity of our approach, we believe our approach will be broadly useful in cell identification, especially when there is a need to identify individual cells of interest. 
This happens when cells are labeled a fluorescent marker, such as in nematode  neuronal or muscle cells.

\section{Compliance with ethical standards}
\label{sec:ethics}

No ethical approval was required for this study.

\section{CODE AVAILABILITY}
\label{sec:code}

The code is publicly available at: \url{https://github.com/pmcesky/4D-Cell}.

\section{Acknowledgments}
\label{sec:acknowledgments}
This work is supported by the Gordon and Betty Moore Foundation under grant number GBMF8815. 
This work was supported by the Howard Hughes Medical Institute. 
This article is subject to HHMI’s Open Access to Publications policy. 
HHMI lab heads have previously granted a nonexclusive CC BY 4.0 license to the public and a sublicensable license to HHMI in their research articles. 
Pursuant to those licenses, the author-accepted manuscript of this article can be made freely available under a CC BY 4.0 license immediately upon publication.

\bibliographystyle{IEEEbib}
\bibliography{4D_cell}

\end{document}